\documentclass{article}
\usepackage[table]{xcolor}
\usepackage{spconf,amsmath,amssymb,graphicx}
\usepackage{booktabs,array,tabularx,capt-of,balance}
\usepackage{tikz,pgfplots}
\usepackage{microtype}
\usepackage[hidelinks]{hyperref}
\pgfplotsset{compat=1.18}

\hypersetup{pdftitle={EvoAudio: Recursive Self-Improvement for Audio Understanding}}

\newcommand{\best}[1]{\textbf{#1}}
\newcolumntype{N}{>{\centering\arraybackslash}p{3.3em}}
\newcolumntype{A}{>{\centering\arraybackslash}p{4.2em}}
\newcommand{\paperfont}{\fontsize{9}{10.8}\selectfont}
\definecolor{rowgray}{gray}{0.92}
\definecolor{headgray}{gray}{0.85}
\definecolor{tblNavy}{HTML}{243B53}
\definecolor{tblBand}{HTML}{EEF2F5}
\definecolor{tblStripe}{HTML}{F6F8FA}
\definecolor{tblOurA}{HTML}{E9F4F1}
\definecolor{tblOurB}{HTML}{EBF6F2}
\definecolor{tblOurInk}{HTML}{174E45}
\definecolor{tblRule}{HTML}{52687A}
\definecolor{tblGain}{HTML}{157951}
\newcommand{\second}[1]{\underline{#1}}
\newcommand{\hd}[1]{\textcolor{tblNavy}{\textbf{#1}}}
\newcommand{\grp}[1]{\textcolor{tblNavy}{\textit{#1}}}
\newcommand{\ours}[1]{\textcolor{tblOurInk}{\textbf{#1}}}
\newcommand{\gain}[1]{{\fontsize{7.2}{7.2}\selectfont\color{tblGain}$+$#1}}
\definecolor{qwen}{RGB}{30,105,170}
\definecolor{mimo}{RGB}{185,70,60}
\definecolor{kimi}{RGB}{20,125,105}
\definecolor{pooled}{RGB}{115,100,65}
\pgfplotsset{paperaxis/.style={
  width=0.475\textwidth,height=3.3cm,
  label style={font=\paperfont},tick label style={font=\paperfont},
  legend style={font=\paperfont,draw=none,fill=white,cells={anchor=west}},
  axis line style={black!65},tick style={black!65},
  grid=major,grid style={black!10},
  every axis plot/.append style={line width=1pt,mark size=1.8pt},
  scaled ticks=false}}

\usepackage{fancyhdr}
\newcommand{\headlogo}[2]{\smash{\includegraphics[height=#1]{#2}}}
\fancypagestyle{preprintfirst}{%
  \fancyhf{}%
  \fancyhead[L]{\headlogo{20pt}{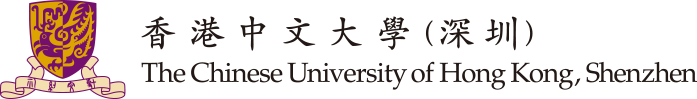}}%
  \fancyhead[C]{\headlogo{16pt}{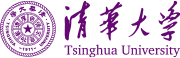}}%
  \fancyhead[R]{\headlogo{13pt}{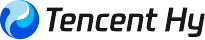}}%
  \fancyfoot[C]{\thepage}%
}
\fancypagestyle{preprintrest}{%
  \fancyhf{}%
  \fancyfoot[C]{\thepage}%
}

\title{EvoAudio: Recursive Self-Improvement for Audio Understanding}
\name{\begin{tabular}[t]{@{}c@{}}
Yuxiang Wang\textsuperscript{*,1,3}, Shengbo Cai\textsuperscript{*,2,3}, Yingda Shen\textsuperscript{1}, Ming-Hao Hsu\textsuperscript{1}, Qinke Ni\textsuperscript{1}\\[2pt]
Liqiang Zhang\textsuperscript{3}, Teddy Sun\textsuperscript{3}, Steve Yevs\textsuperscript{3}, Zhizheng Wu\textsuperscript{1,4}\thanks{\llap{\textsuperscript{*}}Equal contribution.}
\end{tabular}}
\address{\textsuperscript{1}The Chinese University of Hong Kong, Shenzhen\quad%
\textsuperscript{2}Tsinghua University\\%
\textsuperscript{3}Tencent Hunyuan\quad \textsuperscript{4}Amphion Technology Co., Ltd.}

\begin{document}
\ninept
\maketitle
\pagestyle{preprintrest}
\thispagestyle{preprintfirst}

\begin{abstract}
Audio language models understand what is said far better than how it sounds. Closing this gap takes more than data. Detailed acoustic annotation is costly, labels from stronger models inherit their errors and limits, and fixed data cannot adapt as the learner improves. We therefore propose \textbf{EvoAudio}\footnote{Demo page: \url{https://jensenyx.github.io/EvoAudio.github.io/}}, a recursive self-improvement system for audio understanding. To our knowledge, it is the
first to evolve the model, waveforms, questions, and difficulty in one closed loop. EvoAudio uses the current model's performance to set the focus and difficulty of the next training data. A library of audio tools then constructs questions whose answers follow from how the audio was made, providing verifiable supervision without new human annotation. Reinforcement learning updates the model, and validation decides whether it enters the next evolution round. Across 13 rounds, EvoAudio improves five models with different audio encoders and language backbones on MMSU, MMAU-Pro, and MMAR. It achieves the highest average for every backbone, raising overall performance by up to 6.3 points. The improvement unfolds over successive rounds, with each stronger model starting the next round. 
\end{abstract}


\section{Introduction}
\label{sec:intro}
\vspace{-5pt}

Large audio language models (LALMs) answer questions about speech, sound, and music through
one interface~\cite{qwen25,mimo,kimi}, yet their abilities are strikingly
uneven. A model that transcribes a sentence perfectly may still fail to say
which of two speakers has the higher voice, how many people took part in a
conversation, or whether a melody sped up. Recent benchmarks test acoustic
perception and semantic reasoning separately. Their results show that models
understand meaning much better than they perceive acoustic
details~\cite{mmsu,mmaupro,mmar}. One likely reason is that many training
questions ask what was said or what an event means, which can often be
answered from a transcript alone~\cite{omnir1,voxsafe,voxprivacy}, rather than requiring
judgments of acoustic cues.

The obvious remedy is more data of the missing kind, but useful data is hard
to create for two reasons. First, acoustic questions require answers that
match subtle cues in the waveform, and obtaining such labels from humans is
slow and expensive. A stronger LALM such as Gemini can generate tasks and
answers at scale, but this may introduce labeling errors, and the resulting
supervision remains limited by the labeling model's own capabilities. Second,
even accurate labels are not enough unless the data suits the current learner. Questions
that are too easy or too hard provide little training signal, while an
imbalanced task mix can improve one skill at the expense of others. Since
strengths and weaknesses differ across models and change during training,
useful data should be both verifiable and adaptive.

We therefore propose \textbf{EvoAudio}, a recursive self-improvement system
for audio understanding. Once initialized, it lets each LALM shape the
curriculum for its successor without any human intervention. To our knowledge,
it is the first audio system to evolve the model, waveforms, questions, and
difficulty in one closed loop. Each round starts from the current LALM as the
solver. An LLM proposer reads how well the solver handles each skill, then
decides which skills it should practice next and whether their questions
should get easier or harder. EvoAudio then calls audio tools to build the
waveforms these questions ask about. Every answer follows from how the audio
was made or from labels the source recording already carries, so each question
is verifiable without new annotation, and a separate measurement confirms that
the rendered audio still carries the cue the question asks about.
GRPO~\cite{grpo} trains a candidate on them, and a fixed held out set decides
whether it becomes the next solver.

Recent work increasingly recognizes that LALM training should evolve with the
model, although current methods provide only partial adaptation rather than
recursive self-improvement. AudioSkills organizes synthetic audio QA by skill
and difficulty, but settles both before training~\cite{af2}, and SARI orders a
generated QA set from easy to hard~\cite{sari}. AudioRubrics adjusts
reward rubrics as model responses change~\cite{audiorubrics}, while AQA-TTRL
derives pseudo labels from consensus among repeated test time
answers~\cite{aqattrl}, and SI-SDA scores pseudo labels by decoding statistics
to adapt to a new domain~\cite{sisda}. Audio-Zero turns unlabeled contrast
pairs into a verifiable odd listener game~\cite{audiozero}, but both remain
fixed throughout training. Outside audio, Absolute Zero
proposes and solves its own code tasks under an executor that verifies
them~\cite{azr}. Audio has no such executor, so EvoAudio builds the
evidence into the waveform and checks it after rendering.

EvoAudio turns this idea into a broad system with 24 tools covering 47
verifiable question types across 6 families, and it is effective across five
LALMs with different audio encoders and LLM backbones. On MMSU, MMAU-Pro, and
MMAR, it gives every backbone its highest average, with gains of up to 6.3
points, and the weakest acoustic abilities improve the most. These gains
accumulate over 13 evolution rounds, since each improved model exposes a new
set of weaknesses for its successor to train on.
.

\begin{figure*}[t]
\centering
\includegraphics[width=\textwidth]{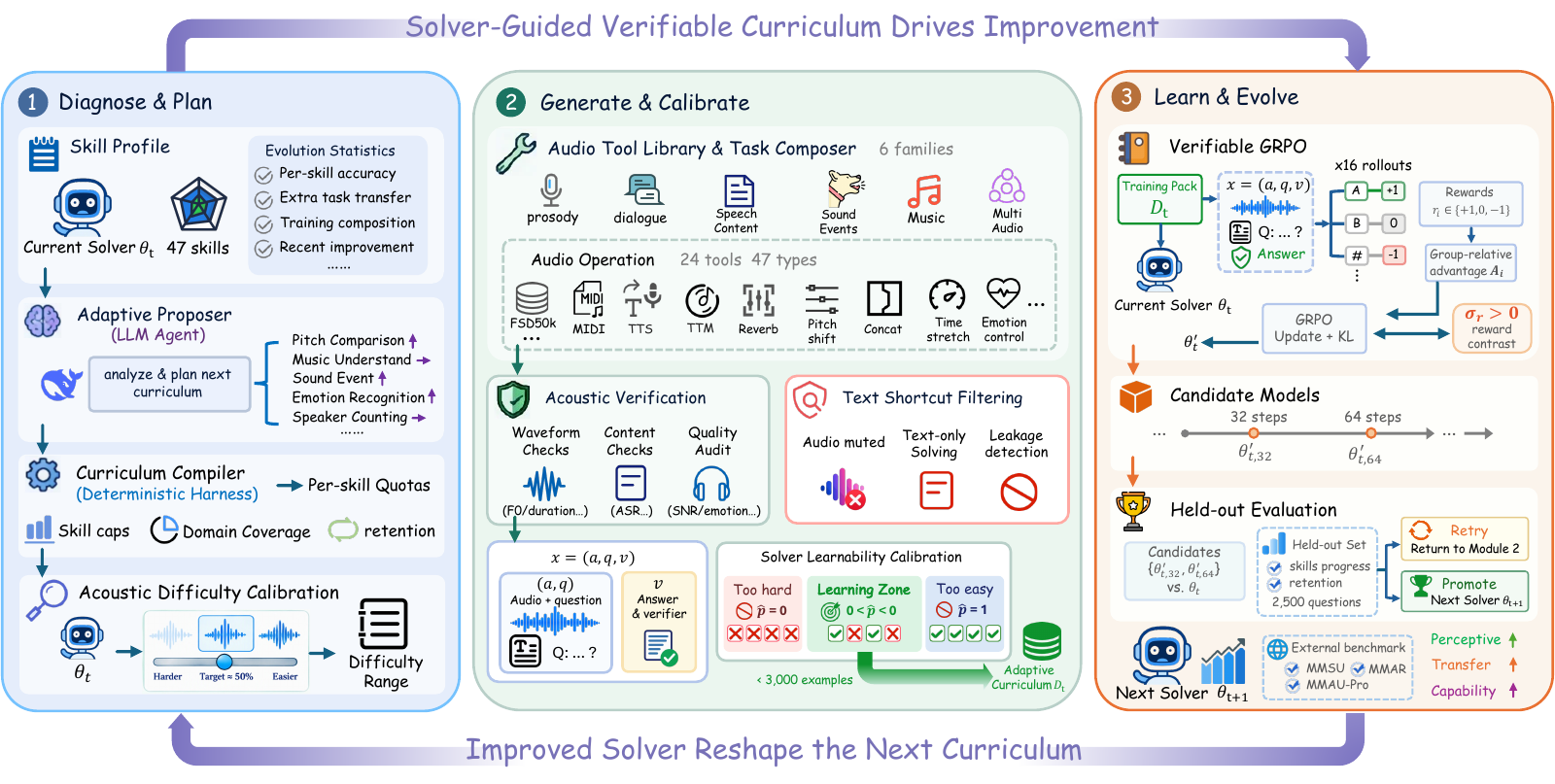}
\par\vspace*{-11pt}
\caption{Overview of the EvoAudio system.}
\label{fig:framework}
\vspace{-5pt}
\end{figure*}

\begin{table*}[t]
\caption{EvoAudio composes 24 tools into 47 verifiable question types across
6 families. Parentheses show type counts.}
\label{tab:coverage}
\centering
\arrayrulecolor{tblRule}
\paperfont
\setlength{\tabcolsep}{3.3pt}
\renewcommand{\arraystretch}{1.0}
\begin{tabularx}{\textwidth}{>{\raggedright\arraybackslash}p{2.65cm}
  >{\raggedright\arraybackslash\hsize=0.95\hsize}X
  >{\raggedright\arraybackslash\hsize=1.15\hsize}X
  >{\raggedright\arraybackslash\hsize=0.90\hsize}X}
\toprule
\textcolor{tblNavy}{\textbf{Family}}
  & \textcolor{tblNavy}{\textbf{What is tested}}
  & \textcolor{tblNavy}{\textbf{How audio is built}}
  & \textcolor{tblNavy}{\textbf{Evidence and checks}} \\
\midrule
\rowcolor{tblStripe}
\textbf{Speech prosody} (9)
  & rate, pitch, loudness, intonation,\newline pauses, syllables, disfluency
  & Qwen3-TTS, time stretch, pitch shift,\newline gain, silence insertion
  & transform record, rate, F0,\newline level, pause, syllable checks \\
\textbf{Speakers and dialogue} (7)
  & identity, similarity, speaker and turn\newline counts, gender, language
  & voice selection, multilingual TTS,\newline emotion control, trimming, concatenation
  & voice IDs, turn plan, counts,\newline gender and language metadata \\
\rowcolor{tblStripe}
\textbf{Speech content} (4)
  & speech QA, intent, emotion
  & scripted TTS, LibriSpeech, MELD
  & scripts, labels, Whisper ASR \\
\textbf{Sound events} (12)
  & count, order, presence, mixtures,\newline level, duration, cause, context
  & FSD50K, AudioSet, procedural synthesis,\newline overlay, mixing, filtering, gain control
  & timeline, gains, source labels,\newline onset and loudness checks \\
\rowcolor{tblStripe}
\textbf{Music} (10)
  & tempo, pitch, dynamics, repetition,\newline instruments, genre, vocals
  & MIDI, FluidSynth, LeVo, stretch,\newline pitch shift, gain ramps, repetition
  & MIDI, transforms, source labels,\newline tempo, pitch, loudness checks \\
\textbf{Multi clip and}\newline\textbf{long scene} (5)
  & cross clip retrieval, sparse event\newline localization, long sequence QA
  & concatenation, long scene assembly,\newline reverb, codecs
  & placement, timeline, transcript,\newline clip and event indices \\
\bottomrule
\end{tabularx}
\arrayrulecolor{black}
\vspace{-5pt}
\end{table*}

\begin{table*}[t]
  \centering
  \caption{Accuracy (\%) on three audio understanding benchmarks.
  Avg.\ is the mean of the three All scores. EvoAudio (SFT) and EvoAudio (GRPO)
  update the solver over 13 rounds with SFT and GRPO, respectively. Both GRPO
  baselines match EvoAudio's optimization budget and training settings and
  differ only in where their questions come from. Static-profile GRPO builds
  fresh items every round and still filters them with the current solver, but
  the base model supplies the skill profile, the quotas, and the difficulty
  settings for all rounds. Pooled GRPO makes a single pass over the union
  of the 13 curricula, split into the same 13 rounds.
  For MMAU-Pro, the three category columns do not cover every scored item, so
  All is not their average. Arrows indicate higher is better, \textbf{bold} and \second{underline} mark
  the best and second best result per model.}
  \label{tab:main}
  \centering
  \fontsize{9}{9.4}\selectfont
  \setlength{\tabcolsep}{1.5pt}
  \renewcommand{\arraystretch}{0.9}
  \arrayrulecolor{tblRule}
  \begin{tabular*}{\textwidth}{@{\extracolsep{\fill}}l*{12}{N}A@{}}
  \toprule
  & \multicolumn{3}{c}{\hd{MMSU $\uparrow$}} & \multicolumn{4}{c}{\hd{MMAU-Pro $\uparrow$}}
  & \multicolumn{5}{c}{\hd{MMAR $\uparrow$}} & \\
  \cmidrule(lr){2-4}\cmidrule(lr){5-8}\cmidrule(lr){9-13}
  \hd{Method} & \hd{Percep.} & \hd{Reason.} & \hd{All} & \hd{Speech} & \hd{Sound} & \hd{Music} & \hd{All} & \hd{Signal} & \hd{Percep.} & \hd{Seman.} & \hd{Culture} & \hd{All} & \hd{Avg. $\uparrow$} \\
  \midrule
  \rowcolor{tblBand}
  \multicolumn{14}{@{}l}{\grp{Qwen2.5-Omni-7B}} \\
  \addlinespace[1.3pt]
  Base & 44.5 & 79.4 & 61.4 & 57.8 & 45.3 & 64.0 & 56.4 & 55.8 & 54.0 & 67.2 & 58.9 & 60.2 & 59.3\phantom{\gain{0.0}} \\
  Static-profile GRPO & 54.7 & \second{81.9} & 67.9 & 61.4 & \best{51.6} & \second{65.9} & 59.5 & \second{58.4} & \second{56.8} & \second{67.3} & \second{62.3} & 61.1 & 62.8\gain{3.5} \\
  Pooled GRPO & 52.0 & 81.1 & 66.1 & 61.7 & 50.5 & 64.7 & 59.2 & 56.9 & 55.9 & 66.6 & 61.5 & 61.0 & 62.1\gain{2.8} \\
  EvoAudio (SFT) & \second{58.5} & \best{82.9} & \second{70.3} & \second{62.0} & 50.9 & 65.2 & \second{59.6} & 55.8 & 56.4 & 67.2 & 61.0 & \second{61.5} & \second{63.8}\gain{4.5} \\
  \rowcolor{tblOurB}
  \ours{EvoAudio (GRPO)} & \best{61.5} & \best{82.9} & \best{71.9} & \best{62.9} & \second{51.3} & \best{67.1} & \best{60.8} & \best{61.6} & \best{57.6} & \best{67.4} & \best{69.4} & \best{63.5} & \best{65.4}\gain{6.1} \\
  \midrule
  \rowcolor{tblBand}
  \multicolumn{14}{@{}l}{\grp{Audio Flamingo 3 (7B)}} \\
  \addlinespace[1.3pt]
  Base & 50.5 & 78.1 & 63.9 & 56.2 & 47.4 & 63.1 & 52.4 & 57.6 & 53.4 & 65.6 & 55.3 & 58.9 & 58.4\phantom{\gain{0.0}} \\
  Static-profile GRPO & 54.1 & \second{79.9} & 66.6 & 59.1 & 52.1 & \second{65.9} & 54.9 & 60.6 & 54.4 & 66.7 & 56.7 & 59.4 & 60.3\gain{1.9} \\
  Pooled GRPO & 52.1 & 79.1 & 65.2 & 58.2 & 49.8 & 64.1 & 54.0 & 59.5 & 53.5 & 65.7 & \second{57.8} & 59.5 & 59.6\gain{1.2} \\
  EvoAudio (SFT) & \second{56.0} & \best{81.4} & \second{68.3} & \second{59.3} & \second{53.9} & 65.6 & \second{55.4} & \second{63.5} & \second{54.6} & \second{67.1} & \second{57.8} & \second{59.6} & \second{61.1}\gain{2.7} \\
  \rowcolor{tblOurB}
  \ours{EvoAudio (GRPO)} & \best{58.0} & \best{81.4} & \best{69.3} & \best{59.4} & \best{56.1} & \best{66.0} & \best{56.9} & \best{69.8} & \best{55.2} & \best{68.7} & \best{58.9} & \best{61.9} & \best{62.7}\gain{4.3} \\
  \midrule
  \rowcolor{tblBand}
  \multicolumn{14}{@{}l}{\grp{Kimi-Audio-7B-Instruct}} \\
  \addlinespace[1.3pt]
  Base & 39.0 & 74.3 & 56.0 & 58.9 & 37.1 & 56.1 & 49.6 & 46.5 & 47.3 & 63.1 & 51.1 & 54.3 & 53.3\phantom{\gain{0.0}} \\
  Static-profile GRPO & \second{44.6} & \second{76.9} & \second{60.2} & \best{63.8} & \second{51.0} & 59.5 & \second{56.8} & \second{54.5} & \second{50.5} & \second{66.4} & \second{57.1} & \second{57.3} & \second{58.1}\gain{4.8} \\
  Pooled GRPO & 42.1 & 75.6 & 58.2 & 61.3 & 45.9 & \second{60.4} & 53.4 & 50.9 & 49.3 & 64.4 & 56.6 & 55.9 & 55.8\gain{2.5} \\
  EvoAudio (SFT) & 43.3 & 76.3 & 59.3 & \second{62.5} & 47.6 & 58.7 & 54.2 & 52.3 & 50.4 & 64.9 & 55.5 & 56.5 & 56.7\gain{3.4} \\
  \rowcolor{tblOurB}
  \ours{EvoAudio (GRPO)} & \best{45.7} & \best{77.3} & \best{61.0} & \best{63.8} & \best{56.9} & \best{60.6} & \best{58.9} & \best{57.0} & \best{51.0} & \best{66.8} & \best{57.6} & \best{58.8} & \best{59.6}\gain{6.3} \\
  \midrule
  \rowcolor{tblBand}
  \multicolumn{14}{@{}l}{\grp{MiMo-Audio-7B-Instruct}} \\
  \addlinespace[1.3pt]
  Base & 46.9 & 73.1 & 59.6 & 62.6 & 38.3 & 64.3 & 54.9 & 44.2 & 55.9 & 69.2 & 62.4 & 61.8 & 58.8\phantom{\gain{0.0}} \\
  \rowcolor{tblOurB}
  \ours{EvoAudio (GRPO)} & \best{52.9} & \best{77.5} & \best{64.9} & \best{65.3} & \best{51.6} & \best{67.6} & \best{61.3} & \best{58.0} & \best{61.2} & \best{72.0} & \best{66.6} & \best{66.4} & \best{64.2}\gain{5.4} \\
  \midrule
  \rowcolor{tblBand}
  \multicolumn{14}{@{}l}{\grp{MiniCPM-o 4.5 (9B)}} \\
  \addlinespace[1.3pt]
  Base & 55.2 & 81.2 & 67.8 & \best{67.7} & 50.1 & \best{67.4} & 60.4 & 53.5 & \best{63.4} & 69.7 & 61.7 & \best{65.3} & 64.5\phantom{\gain{0.0}} \\
  \rowcolor{tblOurB}
  \ours{EvoAudio (GRPO)} & \best{61.8} & \best{82.3} & \best{71.8} & 67.5 & \best{52.4} & 67.2 & \best{61.0} & \best{58.4} & 60.0 & \best{70.8} & \best{62.6} & 65.0 & \best{65.9}\gain{1.4} \\
  \bottomrule
  \end{tabular*}
  \arrayrulecolor{black}
  \vspace{-5pt}
  \end{table*}

\vspace{-5pt}
\section{Recursive Self-Improvement with EvoAudio}
\label{sec:method}
\vspace{-5pt}

EvoAudio maintains a lineage $\theta_0,\theta_1,\ldots$. Round $t$ builds a
verifiable curriculum $\mathcal{C}_t$ matched to the solver
$\pi_{\theta_t}$, trains a candidate $\theta_t'$, and sets
$\theta_{t+1}=\theta_t'$ only after held out
validation~(Fig.\ref{fig:framework}).

\vspace{-3pt}
\subsection{A Curriculum Matched to the Solver}
\vspace{-3pt}

Before generating any audio, EvoAudio decides what the current solver should
practice. It profiles the solver over 47 skills, one per question type. The
accuracy $p_{t,k}$ of skill $k$ and its change
$\Delta_{t,k}=p_{t,k}-p_{t-1,k}$ are read from the held out set described
below, which is fixed across rounds, so they stay comparable even as the
training difficulty moves. The mixed rate $m_{t,k}$, the share of last round's
items whose 16 sampled responses were not all scored alike, comes from
training. The harness also correlates $\Delta_{t,k}$ with the changes on the
other skills, a coarse indicator of whether a skill still moves anything
besides itself. A DeepSeek-V4-Flash proposer~\cite{deepseek} reads this
evidence, returns a priority $\rho_{t,k}$ per skill, and recommends whether
its questions should get easier or harder.

A priority alone ignores whether a skill can currently be trained, so the
harness reweights it by the mixed rate, as in automatic curricula driven by
learning progress~\cite{tscl}, then caps and renormalizes:
\begin{equation}
q_{t,k}\;\propto\;\min\!\left(\rho_{t,k}\,m_{t,k},\;0.06\right),
\qquad \textstyle\sum_{k} q_{t,k}=1 .
\end{equation}
The cap keeps one skill from taking over a round, 12\% of the budget stays
with mastered skills to reduce forgetting, and a skill whose $\Delta$ has
stalled and no longer tracks any other skill gives up quota to weaknesses that
are still moving. Where a skill has a continuous acoustic control $c$, such as
a pitch interval or a signal to noise ratio, the easier or harder
recommendation becomes a step toward a target success rate $p^\star=0.5$
estimated from pilot responses $\hat{p}_{t,k}$ of the current solver,
\begin{equation}
c_{t+1,k}=\mathrm{clip}\!\left(c_{t,k}+\eta\,(\hat{p}_{t,k}-p^\star),\,c_{\min},\,c_{\max}\right),
\end{equation}
where the bounds keep $c$ inside the range the tools render reliably and a
fixed share of each skill holds the previous, easier setting.

\subsection{Constructing Verifiable Questions}

The harness then turns the curriculum into concrete training items. Each item
is a triple $x=(a,q,v)$, where $a$ is the waveform, $q$ is the question, and
$v(y)$ checks a response $y$. Following the plan, EvoAudio builds $a$ from
speech and song synthesis~\cite{tts,levo}, procedural mixing, and labeled
recordings, in the spirit of soundscape synthesis that emits its own
annotation~\cite{scaper}; Table~\ref{tab:coverage} summarizes the library. A
pitch comparison item, for example, renders one sentence with Qwen3-TTS,
resamples a copy by a drawn ratio $r$, and concatenates the two, so which
voice is higher is fixed by the sign of $\log r$ and by nothing else.
Every operation records its parameters, and recordings from LibriSpeech,
FSD50K, AudioSet, or MELD keep the labels they already
carry~\cite{librispeech,fsd50k,audioset,meld}. The answer therefore comes from
construction or from an existing label, never from a model annotation.

Knowing the answer from construction is not enough, because synthesis and
mixing can blur the intended cue, so the two halves of verification are kept
apart. The verifier $v$ only decides whether a response matches the recorded
answer; a separate acoustic check decides whether that answer is still
faithful to the rendered waveform. EvoAudio remeasures pitch, rate, loudness,
onsets, and tempo, discards items whose cue is too weak to hear reliably, and
uses Whisper~\cite{whisper} to check synthesized speech against its script.
For the types that inherit a label, such as emotion or genre, this check
confirms that the transform preserved it. A last probe answers each question
from its text alone and drops the ones that never needed the
audio~\cite{omnir1}.

Validity alone does not make a question useful for the current solver. The
harness runs four solver rollouts on every surviving candidate and keeps those
answered correctly in one to three of them, since a group whose rewards are
all identical provides no GRPO signal~\cite{dapo}; the survivors form
$\mathcal{C}_t$. Validation instead uses a fixed held out set of 2,500
questions, reused in every round, that shares no template, source recording,
or generation seed with any training item.

\vspace{-3pt}
\subsection{Training and Promotion}
\vspace{-3pt}

GRPO trains on $\mathcal{C}_t$. For each item $x=(a,q,v)$, the verifier $v(y)$
assigns rewards $1$, $0$, and $-1$ to correct, wrong, and malformed responses,
and GRPO normalizes them across the responses to that item. A round saves two
candidates, at 32 and 64 steps, and the held out set, which receives no
gradient, scores both against the parent $\theta_t$. The better candidate
becomes $\theta_{t+1}$ if it outscores $\theta_t$, and its skill profile
drives the next round. If neither does, both are dropped and $\theta_t$
remains the solver, so the next round builds a new curriculum from the same
parent. The improved solver therefore shapes the curriculum for its
successor, closing the recursive loop.

\vspace{-5pt}
\section{Experiments}
\label{sec:experiments}
\vspace{-5pt}

\vspace{-3pt}
\subsection{Setup}
\vspace{-3pt}

A recursive improvement method should transfer across architectures rather
than exploit the quirks of one backbone. We therefore study
Qwen2.5-Omni~\cite{qwen25}, MiMo-Audio~\cite{mimo},
Kimi-Audio~\cite{kimi}, MiniCPM-o 4.5~\cite{minicpm}, and
Audio Flamingo 3~\cite{af3}, which combine different audio encoders and
language backbones. For every model and every method, we train all
parameters on the audio and language path and freeze the visual modules and the
unused output branches.

A fair comparison also requires the same opportunity to improve. Every run is
organized into 13 rounds of 64 steps with 16 prompts per step and 16 responses
per prompt, and both GRPO baselines receive that same budget. EvoAudio (SFT)
replaces only the update rule and keeps the curricula, rounds, steps, and
learning rate of EvoAudio (GRPO), supervising each prompt with the answer its
construction recorded. Every run uses a constant learning rate of
$1\times10^{-6}$ and starts each round from a fresh optimizer state; the GRPO
runs use temperature 0.9 and a KL coefficient of 0.02 against the parent.
Table~\ref{tab:main} reports the final promoted model of each run. Training
uses 4 nodes with 8 H800 GPUs each.

No single benchmark covers every dimension of audio understanding. MMSU
tests spoken perception and reasoning~\cite{mmsu}, MMAU-Pro covers speech,
sound, and music~\cite{mmaupro}, and MMAR spans signal, perceptual, semantic,
and cultural reasoning~\cite{mmar}.
For MMAU-Pro, we exclude open ended, spatial, and instruction following
subsets. This leaves 4,268 closed nonspatial items, of which 4,239 have
official answers and are scored. Invalid predictions count as incorrect.

\begin{figure}[t]
\centering
\includegraphics[width=\columnwidth]{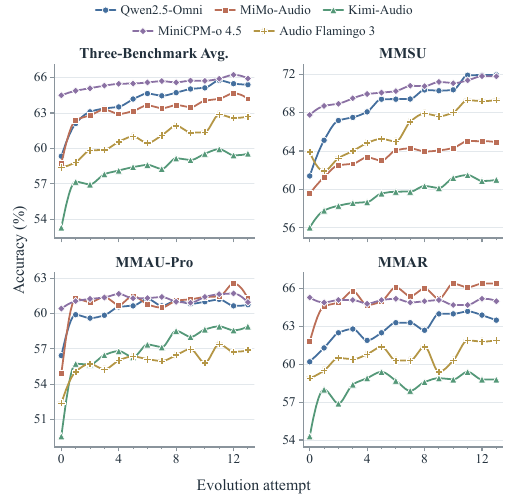}
\par\vspace*{-11pt}
\caption{Self evolution of five backbones. Attempt 0 is the base model and
each later attempt is the candidate that round produced. The top left panel
is the mean accuracy over the three suites.}
\label{fig:curves}
\vspace{-10pt}
\end{figure}

\subsection{Main Results}

Table~\ref{tab:main} shows a consistent ordering. EvoAudio (GRPO) achieves the
highest overall average on every backbone, and it leads every baseline on the
three that carry the full comparison. Relative to Base, its gains span nearly
every benchmark and category, with the few regressions confined to MiniCPM.
On the first three backbones with both variants, EvoAudio (GRPO) also
outperforms EvoAudio (SFT). One likely reason is that construction fixes the
answer but not the reasoning, so the SFT target is a verified answer with no
trace attached, while GRPO reinforces whichever of the model's own attempts
reaches it.
Across these backbones, Pooled GRPO and Static-profile GRPO both
improve over Base but stay below EvoAudio, and Static-profile GRPO is
consistently ahead of Pooled GRPO. Static-profile GRPO still filters each
round's candidates with the current solver and differs from EvoAudio only in
that its priorities, quotas, and difficulty settings stay at the base model's,
so the 2.1 points between them isolate what those decisions are worth. A profile read once from the base model keeps asking for weaknesses
the solver has repaired, and its fixed difficulty falls further
behind as the solver improves. The pooled corpus misses in the other direction,
adding questions written in later rounds for stronger solvers that are often
too hard to produce reward variance. In the final round only 30.8\% of pooled
questions get mixed rewards, against 53.8\% for EvoAudio.

\subsection{Evolution Across Models}

Fig.\ref{fig:curves} shows that all five backbones improve as evolution
proceeds. The small fluctuations are expected, since the curve shows every
candidate, including those promotion rejected, and a candidate that gains on
the held out set does not always gain on the benchmarks. MMSU gains the most:
averaged over backbones its accuracy rises by 6.1 points from Base to attempt
13, against 5.0 on MMAU-Pro and 3.0 on MMAR. This ordering follows the reach
of the tool library. Rate, pitch, loudness, and event structure are what the
tools control and what MMSU asks about, while the semantic and cultural
reasoning of MMAR lies mostly beyond what they can build.

For four of the five backbones, the largest increase occurs in the first
attempt, likely because the base model has many broad yet learnable weaknesses
that yield reward variation. Later rounds add progressively less and
the curves are close to flat by the end. Two effects may explain this. First,
the remaining errors are fewer and harder, so a stronger solver answers most
generated questions the same way and leaves less signal per round. Second, questions
the solver never gets right carry no reward contrast, so reinforcement
alone cannot open those skills. An SFT warm start combined with GRPO may
help here, and is our next step for EvoAudio.

\subsection{Ablations}

\begin{table}[t]
\caption{Each row changes one part of the loop on Qwen2.5-Omni-7B and leaves
everything else, including the training budget, as it was. Avg.\ is the mean
accuracy over the three benchmarks, where the base model scores 59.3\%. Var.\ is
the percentage of training questions in the final round whose 16 sampled
training responses are not all scored alike, which is the condition for a
nonzero GRPO advantage.}
\label{tab:ablation}
\centering
\arrayrulecolor{tblRule}
\paperfont
\setlength{\tabcolsep}{4pt}
\renewcommand{\arraystretch}{1.05}
\begin{tabularx}{\columnwidth}{X>{\centering\arraybackslash}p{5.3em}A}
\toprule
\hd{Variant} & \hd{Avg. (\%) $\uparrow$} & \hd{Var.\ (\%)} \\
\midrule
\rowcolor{tblOurB}
\ours{EvoAudio (GRPO)} & 65.4 & 53.8 \\
Difficulty fixed at round one & 63.3 & 40.7 \\
Keep every verified question & 63.6 & 34.5 \\
Uniform skill quotas & 63.7 & 45.3 \\
No retention floor & 64.9 & 52.4 \\
No acoustic verification & 64.1 & 57.2 \\
\bottomrule
\end{tabularx}
\arrayrulecolor{black}
\vspace{-10pt}
\end{table}

Table~\ref{tab:ablation} reveals a clear pattern. Freezing difficulty and
keeping every verified question cause the largest drops in accuracy and also
produce the lowest Var. Both push questions outside the solver's learning
range, where its repeated answers no longer include a useful mix of successes
and failures, which weakens the signal available to GRPO. Uniform skill quotas
cost slightly less, and removing the retention floor changes little over 13
rounds.This suggests that matching difficulty to the current
solver matters more here than precisely allocating skills or reserving
questions for skills it already handles well.

Acoustic verification reveals the other half of the story. Removing it raises
Var.\ from 53.8\% to 57.2\% but lowers accuracy from 65.4\% to 64.1\%, so more
reward variance is not always useful. Without the check, a rendering error can
change the audio while the recorded answer stays the same; such items still
produce mixed rewards, but the rewards now point toward a wrong answer.
EvoAudio therefore needs questions that are both informative for the current
solver and faithful to the rendered audio.

\vspace{-2pt}
\section{Conclusion}
\label{sec:conclusion}
\vspace{-2pt}

We introduced EvoAudio, the first recursive self-improvement system to jointly
evolve the model, waveforms, questions, and difficulty in one closed loop. It
creates verifiable training data and adapts each new curriculum to the improved
model, yielding broad gains across five LALMs. What bounds this loop is not the
supply of labeled audio but the reach of the tools that build its questions,
so richer tools and stronger verifiers are what would widen it.

\clearpage
\balance
\paperfont\sloppy
\bibliographystyle{IEEEbib_etal}
\bibliography{refs}
\end{document}